\documentclass[sigconf]{acmart}

\usepackage{color,soul}
\usepackage{graphicx}

\AtBeginDocument{%
  \providecommand\BibTeX{{%
    \normalfont B\kern-0.5em{\scshape i\kern-0.25em b}\kern-0.8em\TeX}}}

\copyrightyear{2026}
\acmYear{2026}
\setcopyright{cc}
\setcctype{by}
\acmConference[TEI '26]{Twentieth International Conference on Tangible, Embedded, and Embodied Interaction}{March 08--11, 2026}{Chicago, IL, USA}
\acmBooktitle{Twentieth International Conference on Tangible, Embedded, and Embodied Interaction (TEI '26), March 08--11, 2026, Chicago, IL, USA}
\acmPrice{}
\acmDOI{10.1145/3731459.3773331}
\acmISBN{979-8-4007-1868-7/2026/03}

\begin{document}

\title{FluxLab: Creating 3D Printable Shape-Changing Devices with Integrated Deformation Sensing}

\author{Hsuanling Lee}
\email{hsuanlinglee22@gmail.com}
\orcid{0009-0000-4832-0458}
\affiliation{%
  \institution{University of Texas at Dallas}
  \city{Richardson}
  \state{TX}
  \country{USA}
}

\author{Jiakun Yu}
\email{yu1591@purdue.edu}
\orcid{0009-0005-7623-6495}
\affiliation{%
  \institution{Purdue University}
  \city{West Lafayette}
  \state{IN}
  \country{USA}
}

\author{Shurui Zheng}
\email{zheng862@purdue.edu}
\orcid{0009-0000-0111-9503}
\affiliation{%
  \institution{Purdue University}
  \city{West Lafayette}
  \state{IN}
  \country{USA}
}

\author{Te-Yen Wu}
\email{tw23l@fsu.edu}
\orcid{0000-0003-3977-9093}
\affiliation{%
  \institution{Florida State University}
  \city{Tallahassee}
  \state{FL}
  \country{USA}
}

\author{Liang He}
\email{liang.he@utdallas.edu}
\orcid{0000-0003-4826-629X}
\affiliation{%
  \institution{University of Texas at Dallas}
  \city{Richardson}
  \state{TX}
  \country{USA}
}

\renewcommand{\shortauthors}{Lee, et al.}

\newcommand{\teyen}[1]{{\small\textcolor{blue}{\bf [TW: #1]}}}
\newcommand{\liang}[1]{{\small\textcolor{red}{\bf [LH: #1]}}}
\newcommand{\hannah}[1]{{\small\textcolor{green}{\bf [HL: #1]}}}
\newcommand{\jiakun}[1]{{\small\textcolor{purple}{\bf [JY: #1]}}}
\newcommand{\quotes}[1]{``#1''}
\begin{abstract}

We present \textit{FluxLab}, a system comprising interactive tools for creating custom 3D-printable shape-changing devices with integrated deformation sensing. To achieve this, we propose a 3D printable nesting structure, consisting of a central SMA channel for sensing and actuation, lattice-based padding in the middle for structural support and controllable elasticity, and parallel helix-based surface wires that preserve the overall form and provide anchoring struts for guided deformation. We developed a design editor to embed these structures into custom 3D models for printing with elastic silicone resin on a consumer-grade SLA 3D printer and minimal post-printing assembly. A deformation authoring tool was also developed for users to build a machine learning-based classifier that distinguishes desired deformation behaviors using inductive sensing. Finally, we demonstrate the potential of our system through example applications, including a self-deformable steamer bowl clip, a remotely controllable gripper, and an interactive desk lamp.

\end{abstract}



\begin{teaserfigure}
    \centering
    \includegraphics[width=\textwidth]{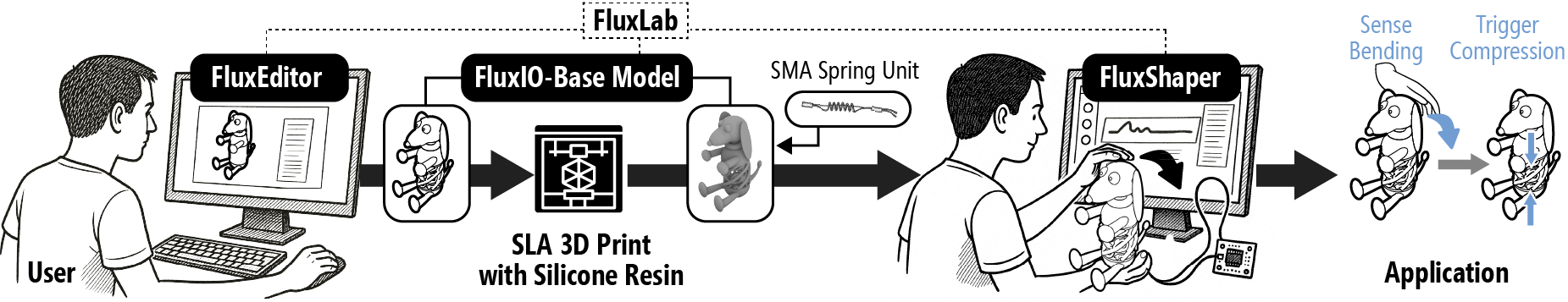}
    \caption{With \textit{FluxLab}, the user first converts the target 3D body into a \textit{FluxIO}-based structure using an interactive design editor---\textit{FluxEditor}. The generated model is then printed on an SLA 3D printer with silicone resin, and the spring SMA and wire are inserted into the body in a post-printing process. Finally, the user trains desired shape-changing behaviors using a machine learning classifier with a deformation authoring tool---\textit{FluxShaper}---which exports reusable auto-generated code in custom applications.}
    \label{fig:teaser}
\end{teaserfigure}

\maketitle

\section{Introduction}
\label{sec:intro}
Shape-changing interfaces have opened up new possibilities for interactive systems that adapt their physical forms in response to user inputs and environments. 
They enable rich, embodied experiences that go beyond the capabilities of traditional static interfaces~\cite{MechSense, MetaSense}.
Recent advances in materials for 3D printing and 3D printable structures have presented a promising approach for making shape-changing systems more accessible with commercially available hardware.
To create 3D printable shape-changing devices, researchers have explored integrating soft materials in 3D-printed objects~\cite{Peng2015, IrOnTex, Forman2020, teddybear2014}, applying environmentally reactive material composition in 3D printing~\cite{A-line, 4DMesh, ShrinCage, Thermorph}, embedding 3D printable kinematic mechanisms~\cite{Ondule, metamaterialIon2016, Codedskeleton, meshMoritz2012, motionZhu2012}, and creating pneumatically controllable 3D-printed bodies~\cite{PneuHaptic, vazquez20153d, Slyper2012, Ma2017}.
Similar to approaches that empower rigid 3D-printed objects with interactivity~\cite{ModElec, Xstrings, RetroFab, Capricate}, sensing and actuation capabilities can be added to 3D-printed shape-changing devices either using conductive 3D printing materials~\cite{MetaSense, DefSense, LattiSense} or interfacing with external sensors and actuators~\cite{MorphSensor, MorphingCircuit, Moon2024, HotFlex}.
Despite these advancements, designing and fabricating such shape-changing systems with integrated sensing and actuation behaviors remains technically demanding, often requiring specialized materials, complex mechanical integration, and tightly coupled sensing-actuation pipelines.

To address these challenges, we introduce \textit{FluxLab}---a system designed to support experienced makers and researchers in creating custom shape-changing devices with integrated deformation sensing capabilities for interactivity. 
With this system, we aim to demonstrate tool support for a complete pipeline of making interactive physical computing devices, encompassing shape-changing mechanical design, fabrication, and integrated sensing control.
At the core of the system is \textit{FluxIO}, a structural design method that defines how to embed sensing and actuation channels directly within 3D-printed forms. The method employs a three-layer 3D-printable architecture (Fig.~\ref{fig:teaser}\&\ref{fig:fluxio_design}): a central channel that houses a shape-memory alloy (SMA) spring serving both as an actuator and a deformation sensor, a lattice padding layer for structural support and controlled flexibility, and a parallel helix-based surface wireframe as an outer layer that preserves the overall shape and selectively scaffolds solid areas for guided deformation, such as lateral bending. 
FluxIO leverages a single SMA component to perform both functions---powering shape change and detecting deformation through inductive sensing---and demonstrates how sensing and actuation can emerge from the same material element and geometric design.

To convert custom 3D models into FluxIO-based shape-changing objects, we developed an interactive design editor, \textit{FluxEditor}, that implements the FluxIO method and enables users to configure the output deformation behaviors. 
The resulting models can be printed on a consumer-grade stereolithography (SLA) 3D printer with an off-the-shelf spring SMA inserted and fixed in the printed object during the post-printing process. 
To enable deformation recognition, we also developed a deformation authoring tool, \textit{FluxShaper}, that takes in the inductive signals when the SMA is connected to an external evaluation board and trains an ad-hoc machine learning-based classifier to recognize various deformation interactions, such as bending and twisting, using inductive sensing (Fig.~\ref{fig:teaser}). 
The built classifier and executable code snippets are exported for custom interactive applications. With FluxLab, we demonstrate several applications, including a self-deformable steamer bowl clip, a remotely controllable gripper, and an interactive desk lamp.
While our work explored the potential of making actuators and sensors in a uniform 3D printable structure, through the specialized SLA 3D printer and material, we envision that FluxLab provides a reproducible and extensible framework for integrating actuation and sensing into 3D-printed forms and will enrich the field of 3D printable interactivity with further investigation and evaluation of alternative accessible mechanisms and materials.

In summary, this paper contributes:  
\begin{enumerate}
    \item \textbf{FluxIO}, a structural design method with a three-layer 3D printable design that can be integrated into 3D models and parameterizable for deformation sensing and actuation;
    \item \textbf{FluxEditor}, an interactive design editor that converts custom 3D models into shape-changing devices with deformation sensing and actuation capabilities by integrating the proposed three-layer structure;  
    \item \textbf{FluxShaper}, a deformation sensing tool for makers to build a machine learning-based classifier to recognize desired deformation behaviors via inductive sensing; and  
    \item A suite of example applications that demonstrate the deformation sensing and shape-changing actuation created with our system.  
\end{enumerate}

\section{Related Work}
\label{sec:rw}
Our work builds on prior research that explores methods for embedding sensing or actuation capabilities into 3D-printed objects. It also relates to approaches for programming and controlling interactive, 3D printable devices.

\subsection{3D Printable Objects with Integrated Sensing Capabilities}
Recent research has explored embedding external sensors~\cite{CurveBoards, FlexBoard} into 3D-printed objects to enable sensing capabilities, such as detecting bending~\cite{salman2020wireless}, pressure~\cite{Yu2024}, acoustic~\cite{squeezapulseHe2017}, and hand gestures~\cite{MyoSpring}. However, these approaches typically require precise placement and manual assembly of external components, making the fabrication process error-prone and time-consuming.
Besides using external sensors, conductive materials have been applied to 3D-printed deformable objects to create 3D printable sensors for sensing user interactions, such as touching or stretching~\cite{Flexibles, Capricate, IrOnTex}. To support deformation sensing with conductive materials, researchers have explored integrating conductive filaments into specialized structures that enable deformation. For example, resistive sensing-based approaches such as \textit{LattiSense}~\cite{LattiSense} and \textit{DefSense}~\cite{DefSense} embed conductive TPU within lattice or internal channels, enabling predictable changes under deformation. Capacitive sensing-based methods, such as \textit{MetaSense}~\cite{MetaSense}, \textit{MorphIO}~\cite{MorphIO}, and \textit{SenSequins}~\cite{SenSequins}, incorporate embedded structures like metamaterial, porous, or sequins structures to produce measurable capacitive variations during deformation. 

Different from prior research, our work focuses on using inductive sensing through coils in a spring SMA to detect deformation behaviors, eliminating the need for precise alignment of embedded electronics or conductive materials through minimal post-printing assembly. Inductive sensing has been shown to offer several advantages over resistive and capacitive methods, including a relatively larger sensing range~\cite{du2014resistive, Laskoski2012}, higher sensing resolution~\cite{IDem, ProjectTasca, Indutivo}, and suitability for use with spring-shaped structures to detect complex deformations~\cite{Lee2024UIST, Lee2024SIGGRAPH, Babkovic2023, Sahu2022}. 

\subsection{3D Printable Actuators}
Previous work on 3D-printed mechanical actuators has explored various methods, including motors~\cite{benditXu2018,RetroFab}, pneumatics~\cite{AirLogic, AirTouch}, and hydraulic systems~\cite{FabHydro,hydraulicsMacCurdy2016}. For example, Ramakers \textit{et al.}~\cite{RetroFab} used standard DC, servo, and stepper motors to achieve the desired mechanical movements. Savage \textit{et al.}~\cite{AirLogic} embedded predefined pneumatic tube-based routes in 3D-printed models to develop inputs and logic for interactive devices. Yan \textit{et al.}~\cite{FabHydro} proposed methods for creating hydraulic actuators with an affordable SLA 3D printer. However, these approaches often require bulky setups and precise control, making it difficult for novices to create portable, accessible, and functional actuators. Shape-memory actuators, including shape-memory polymers (SMPs) and SMAs, offer potential solutions to these challenges. Prior work has demonstrated that both SMPs~\cite{ShrinkCells, EpoMemory} and SMAs~\cite{ClothTiles, Bendi} can return to a pre-deformed shape when heated by embedded circuits, and can be re-deformed upon cooling. By controlling heat input, researchers can control these shape changes. SMAs, in particular, offer higher recovery stress and better fatigue resistance, making them more suitable for long-term and repetitive deformation~\cite{tobushi2009two}.
As a result, SMAs have been increasingly embedded into 3D-printed structures through tailored mechanical designs~\cite{Codedskeleton}. This approach enables the fabrication of compact, self-actuating mechanisms without relying on bulky motors or pneumatic systems, lowering the entry barrier for creating shape-changing devices.

Our work focuses on embedding the spring SMA into 3D-printed lattice structures, serving a dual purpose — enabling both deformation sensing and shape changing, unlike previous approaches that typically support only one of these integrated capacities.

\subsection{Design Tools for 3D Printing Interactivity}
Our work also relates to the body of research that focuses on developing tools to enable end-users to design and fabricate 3D-printed devices with desired behaviors~\cite{trilaterate, ModElec, MetaSense}.
Similar to prior work that converts 3D models into specialized structures~\cite{Ondule, Kinergy, metamaterialIon2016,telescopingMegaro2017,X-Bridges, ThermoFit}, our work comprises an interactive design editor that allows the user can directly edit and preview the generated geometric results in real time. 
Previous work suggests a user-friendly interface for end-users, especially novices, to customize 3D designs through parameterization without knowing the underlying mechanisms. 
For example, \textit{Ondulé}~\cite{Ondule} employs easy-to-control sliders and buttons for the end-user to adjust the stiffness of the embedded spring in 3D models.
Our work applies the concept in the design editor, where the end-user focuses on tuning their high-level intent, such as the overall elasticity of the 3D body, rather than the detailed lattice structures.
In addition to an interactive design editor, our system also provides a deformation authoring tool that enables the end-user to configure desired deformation behaviors of the 3D-printed object.
Inspired by the sensing authoring tools in previous work~\cite{DefSense, EIT-KIT2021Zhu, MetaSense}, our tool guides the end-user through collecting sensor data, labeling deformation behaviors, and training classifiers. This allows the end-user to prepare the optimal deformation sensing toolkit for their custom applications. 
The design editor and deformation authoring tool provide a streamlined pipeline for creating responsive, shape-changing devices, eliminating the need for domain knowledge of mechanical design or deformation sensing.

\section{F\MakeLowercase{lux}IO Design}
\label{sec:sensor_actuator}
\begin{figure}[h]
    \centering
    \includegraphics[width=1\linewidth]{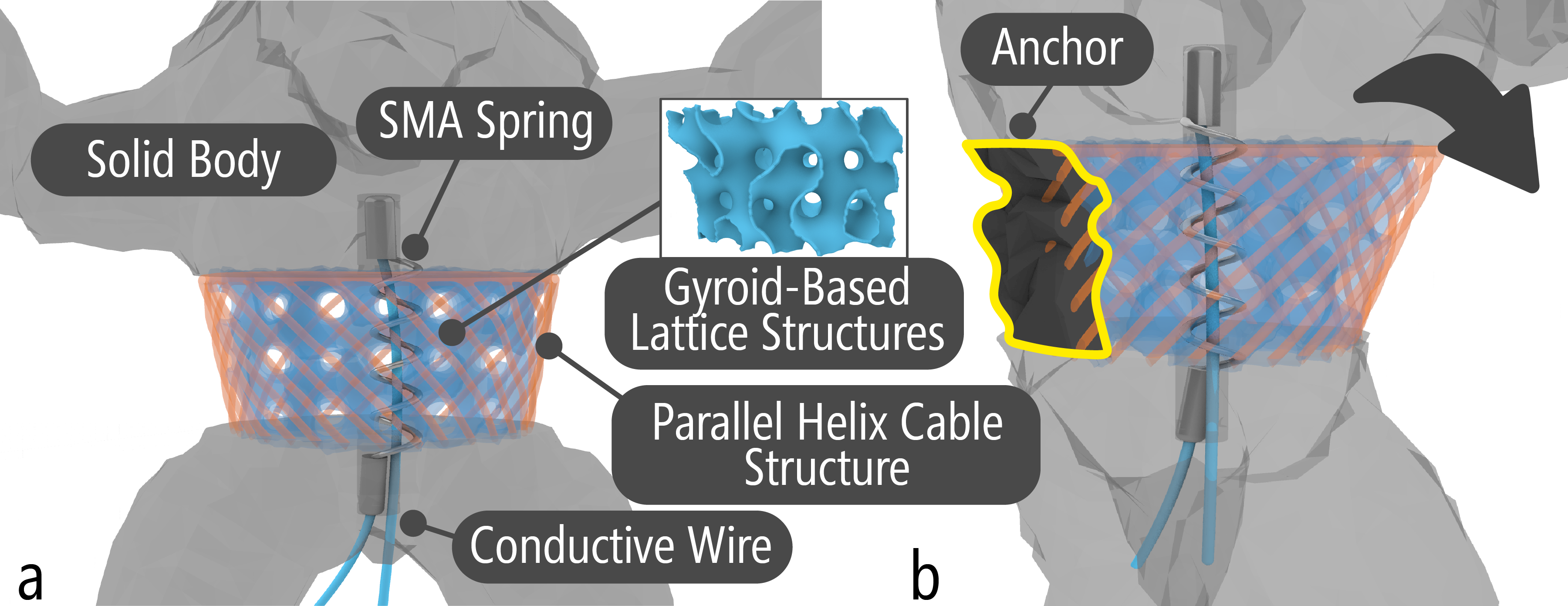}
    \caption{The integrated FluxIO mechanism comprises three components (a): a two-way Nitinol SMA spring (inner), gyroid-based lattice padding (middle), and parallel helix-based surface wireframe (outer). (b) A solid surface anchor can be added to produce lateral bending. }
    \Description{The figure shows the three components in the lattice-based FluxIO structure. }
    \label{fig:fluxio_design}
    \vspace{-2mm}
\end{figure}

To convert a custom 3D shape into a soft, shape-changing part with integrated sensing capabilities, we introduced FluxIO, a lattice-based structure that houses an SMA spring for both sensing and actuating purposes (Fig.~\ref{fig:fluxio_design}a).
FluxIO consists of three key components: an SMA spring (center), lattice-based padding (middle), and surface wireframe (outer).
An off-the-shelf SMA spring is securely housed in the central cylindrical channel as the core element for the input and output capacities: \textit{inductive sensing} when the current goes through the SMA coils and \textit{actuation} when the SMA is heated.
Surrounding the channel, the 3D body is filled with lattice structures to provide homogeneous padding and deformation when the SMA is actuated.
Finally, to approximate the model's appearance and accommodate for deformation, parallel helix-based wires are used to form a wireframe that conforms to the original body shape.
With the lattice structures evenly distributed around the central channel, a default uniform compression is yielded when the SMA is heated.
However, the solid surface regions can be preserved to translate the compression into controllable lateral bending (Fig.~\ref{fig:fluxio_design}b).
Below, we describe each component in detail.

\subsection{SMA-Based Actuation and Sensing}
\begin{figure}[h]
    \centering
    \includegraphics[width=1\linewidth]{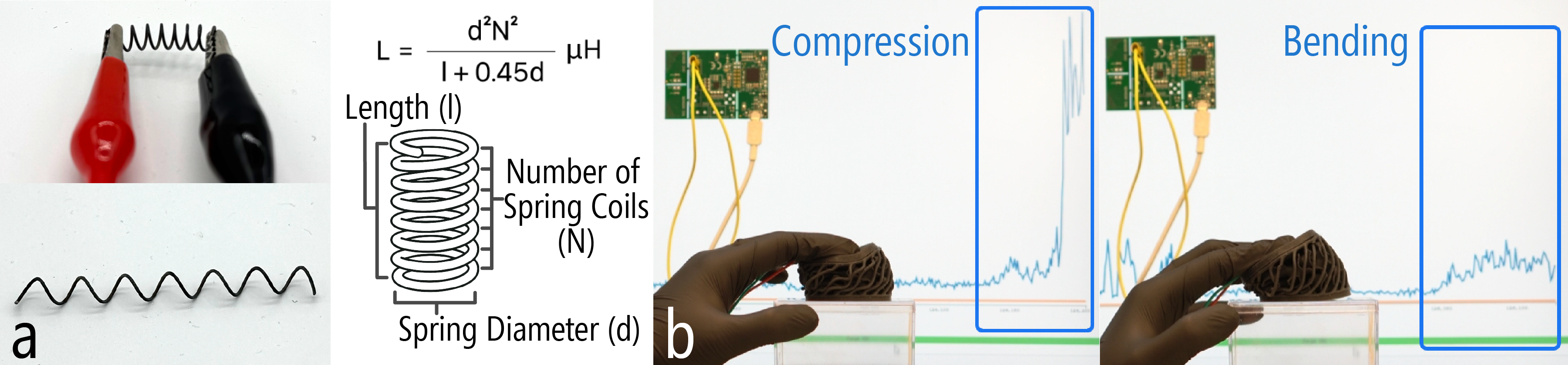}
    \caption{The spring SMA becomes a (a) coil-based inductor for inductive sensing under different body deformation, such as (b) compression and bending.}
    \Description{The figure shows the principles used in our work for inductive sensing }
    \label{fig:sensing}
    \vspace{-2mm}
\end{figure}

A two-way Nitinol SMA spring is embedded into the 3D-printed body to provide sensing and actuation capabilities in a uniform form with a series of coils. The SMA spring contracts when heated and expands and returns to the original state when cooled by itself without any pullback mechanism. As the SMA spring connects model parts, the two connecting parts move closer or apart under temperature control, exhibiting salient shape-changing movements (Fig.~\ref{fig:sensing}b). In the meantime, the SMA becomes a coil-based inductor when the current travels through the coils. When the SMA deforms, the coil-based inductor situated inside the body alters its length, resulting in a change of the magnetic field flux built up by the coils (Fig.~\ref{fig:sensing}a). The inductance signals yielded are recognized as distinct deformation behaviors---\textit{compression, extension, bending, twisting, compression+twisting, extension+twisting}---using a machine learning-based classifier.

\subsection{Lattice-Based Padding for Controllable Shape Changing}
To ensure a uniform shape change along with the actuation of the central embedded SMA spring for control, we added lattice structures as the compliant padding structures around the SMA channel (Fig.~\ref{fig:fluxio_design}a).
The lattice structures offer two benefits: (1) The repeating arrangement of cellular structures converts the solid body into a homogeneous, soft part that performs stable, uniform deformations driven by the SMA actuator.
(2) It is easy to print lattice structures without support on an SLA 3D printer, which is essential in our approach, as additional support could compromise both the mechanical behaviors and the visual aesthetics of the object.
Based on the characteristics reported for various cell types for additive manufacturing \cite{latticereview}, the gyroid-based lattice, a kind of triply periodic minimal surface structure (TPMS), outperforms other lattice forms in terms of homogeneity, strength, and weight.
In addition, based on our empirical tests with lattice structures printed with our selected elastic, silicone-like material, we anecdotally concluded that surface-based lattice (\textit{e.g.,} gyroid, schwarz), compared with strut-based lattice (\textit{e.g.,} cubic, body-centered cubic, tetrahedral), yielded a higher success rate for printing without the supporting material. As a result, we used the gyroid-based lattice, which is isotropic, as the padding layer in our design method.

The gyroid-based lattice structure has a smooth, wavy shape that repeats in all directions without flat surfaces or straight lines, forming two separate yet connected spaces that twist around each other.
When converting a solid body to this unique lattice-based shape, two primary parameters are controlled to design the gyroid-based lattice structure---wall thickness (\textit{T}) and unit cell size (\textit{S}). 
Increasing wall thickness makes the structure stronger and stiffer, while thinner walls may be challenging to fabricate accurately, leading to defects in 3D printing. 
Smaller unit cells distribute loads more evenly and enhance the overall strength and toughness; however, it can be challenging to produce small cells reliably through 3D printing. By controlling the wall thickness and unit cell size with printability considerations, we can create 3D shapes within a range of solidities.

\subsection{Helix-Based Surface Approximation}
To approximate the body's appearance and preserve its aesthetics, we convert the body's surface into a parallel helix-based surface wireframe (Fig.~\ref{fig:fluxio_design}a) that closely conforms to the original organic 3D shape. We choose the parallel helix cable structure because (1) it requires fewer wires to preserve the surface geometry without impacting the overall density of the filled gyroid-based lattice structures, and (2) it is friendly to compression, which is the default shape-changing behavior caused by the contraction of the embedded SMA spring.

\subsection{Anchors for Controllable Deformation}

\begin{figure}[h]
    \centering
    \includegraphics[width=1\linewidth]{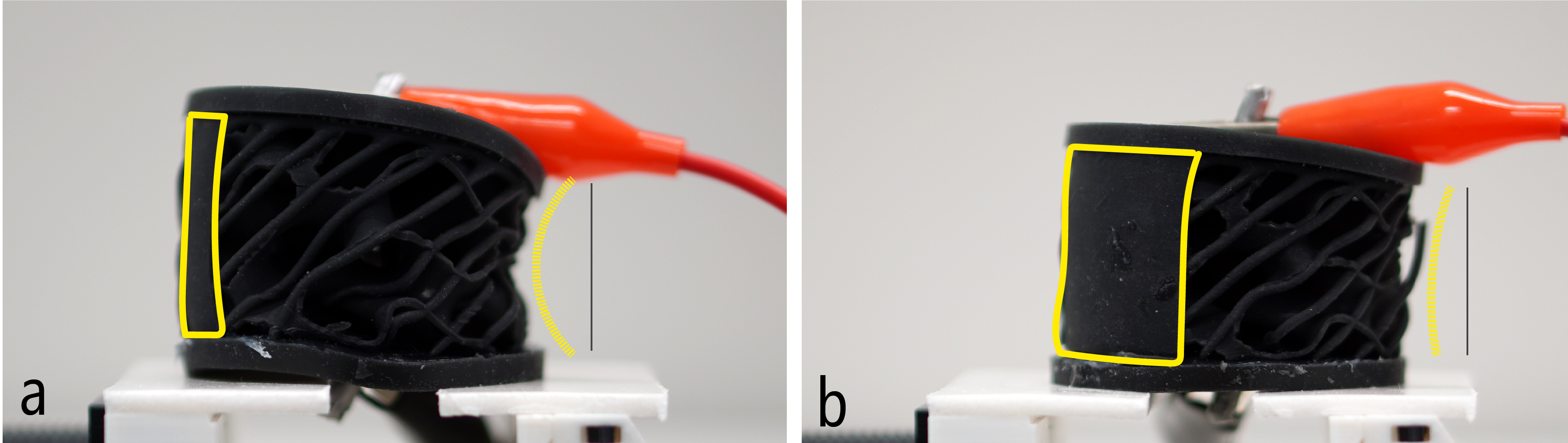}
    \caption{The anchor width is a contributing factor to impact the bending behavior:  (a) a narrow anchor leads to a larger lateral bending angle compared with (b) the one with a wide anchor.} 
    \Description{The experiments to explore the impact of lattice solidity and model size on the compression behavior and anchor experiments for bending behaviors. }
    \label{fig:techeval}
    \vspace{-2mm}
\end{figure}

Besides compression triggered by heating the embedded SMA spring, we also introduce a technique, \quotes{Anchoring}, to translate the uniform compression into controllable lateral bending behaviors (Fig.~\ref{fig:techeval}b\&c).
By selectively preserving one continuous surface region of the body as the solid \quotes{anchor}, we create an asymmetrical material distribution that induces directional bending away from the solid side due to differential compliance. 
Based on empirical tests on the dimensions of the surface anchor, we found that anchor width, rather than anchor thickness, is the contributing factor to bending behavior. For example, compared to a wide anchor (Fig.~\ref{fig:techeval}c), a narrower anchor (Fig.~\ref{fig:techeval}b) more readily induced larger bending angles.
By controlling the anchor position on the perimeter, we can direct the body to perform controllable bending as the central SMA spring contracts under heating.

\section{F\MakeLowercase{lux}L\MakeLowercase{ab} System}
\label{sec:system}
FluxLab comprises two components: an interactive design editor (FluxEditor) and a deformation authoring tool (FluxShaper). 
To streamline the creation of shape-changing devices with embedded sensing, the user first converts rigid 3D models into deformable structures with FluxEditor. The user selects and converts a target region into a FluxIO-based structure, adjusts the overall elasticity of the converted body part, and customizes the desired output deformation behaviors (\textit{i.e.,} bending, compression) through real-time previews. After the model is exported from FluxEditor and printed, the user switches to FluxShaper to program the printed device to recognize deformation inputs (\textit{e.g.,} twisting, extension). To do so, the user follows the interface and performs input gestures on the print for data collection. Once data is collected, the user trains a machine learning-based classifier, which is exported for custom use in applications. With both components, the user can create custom shape-changing devices with configured deformation sensing capabilities without specific expertise in 3D modeling, mechanics, and machine learning.

\begin{figure*}
    \centering
    \includegraphics[width=0.8\linewidth]{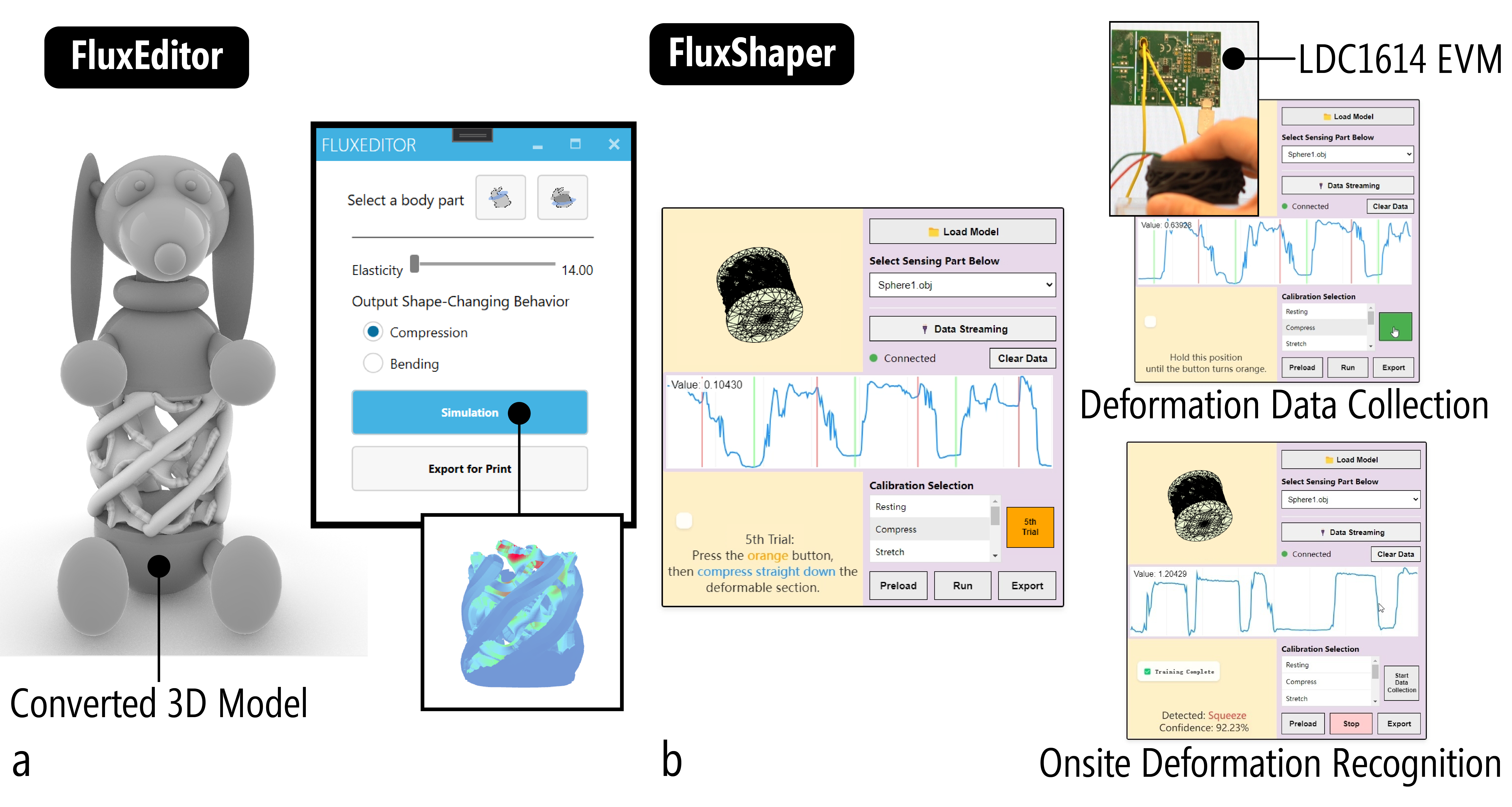}
    \caption{The user interfaces of (a) FluxEditor and (b) FluxShaper.}
    \Description{The figure shows the user interface of FluxEditor and FluxShaper. }
    \label{fig:systems}
    \vspace{-2mm}
\end{figure*}

\subsection{FluxEditor: Converting 3D Models to FluxIO-Based Structures}

FluxEditor aims to support the user, regardless of their 3D modeling expertise, in converting arbitrary regions of a 3D model into deformable, shape-changing FluxIO-based structures. This design editor provides a guided workflow with real-time visual feedback, allowing the user to quickly iterate on 3D designs, adjust body elasticity, configure output shape-changing behaviors, and validate results before fabrication. The user can create a custom FluxIO-based 3D model with desired elasticity and shape-changing behaviors with the following steps. 

\paragraph{Body Part Selection.} Upon launching the editor, the user first selects a region of interest on the 3D model. By dragging two planes that slice the 3D model, the user decides the selected segment, and the body in that segment is automatically converted into a FluxIO-based structure. 

\paragraph{Elasticity Control.} After conversion, the user can adjust the elasticity of the converted body using the \textit{Elasticity} slider (Fig.~\ref{fig:systems}a). When the user drags the slider, the embedded lattice units are automatically updated to map to the selected elasticity. The range of the slider is determined by the lattice parameters empirically optimized for SMA compatibility.

\paragraph{Output Shape-Changing Behavior Selection.} The user then selects a desired output behavior, either \textit{Compression} or \textit{Bending}, with \textit{Compression} set as the default. If \textit{Bending} is selected, the system prompts the user to define the bending direction by asking the user to position the cursor at a location around the model. With the selected direction, anchoring structures on the model surface are automatically generated to constrain and guide the deformation. At each step, a real-time 3D preview is updated to display the generated result, including the lattice density, the surface wireframe, the generated anchors, and the embedded SMA path.

\paragraph{Actuation Preview.} After adjusting the elasticity and configuring the output behaviors, the user can preview the body's actuation response through a built-in simulation. 
The simulation animates the predicted output actuation and uses a color gradient to indicate strain distribution, with warmer colors representing high-strain zones and cooler colors indicating low-strain regions. 
This visual feedback helps users identify critical areas and adjust structure or anchor parameters before finalizing the model for 3D printing.

Finally, the structure is exported in a printable format with embedded channels, lattices, and sockets that are used for SMA installation in the post-printing process, ready for fabrication.

\subsection{FluxShaper: Authoring Deformation Input}

FluxShaper extends FluxEditor's shape-changing behavior customization by enabling deformation input authoring through a three-step workflow coupled with a guiding user interface (Fig.~\ref{fig:systems}b).

\paragraph{Step 1 - Preparation.} The user starts by clicking on the \textit{Load Model} button to import the converted FluxIO-based 3D design. Then, the user selects shape-changing regions (\textit{i.e.,} FluxIO-based parts) from a drop-down list. After connecting the evaluation board to the SMA spring situated in the target 3D-printed FluxIO-based body, the inductive signal is live-streamed in the user interface. The user clicks \textit{Data Streaming} to initiate a five-second normalization phase. 

\paragraph{Step 2 - Training.} After the setup is ready, the user starts the deformation data collection. The user first selects a deformation type---Resting, Compression, Extension (Stretch), Twisting, Bending, Compression \& Twisting, and Extension (Stretch) \& Twisting---from the calibration list, and then clicks on the \textit{Start Data Collection} button to start data collection with each input deformation gesture. For each deformation, the interface guides three repetitions as an orange button and panel message (``Press the orange button, then...'') indicate the beginning of the recording progress. The orange button turns green when clicked and remains green for five seconds for data recording with a message displayed (``Hold this position...''). Sensor streams update in the user interface, with collected sequences stored per deformation type. Clicking the \textit{Run} button initiates training of the classifier, and the training progress is displayed through a dynamically updating report of training and testing accuracy.

\paragraph{Step 3 - Deployment \& Deformation Recognition.} After the training is completed, the user can test the classifier's accuracy by physically manipulating the 3D print with desired deformation inputs, as the predicted deformation input gesture is displayed on the user interface for validation. Finally, upon clicking on the \textit{Export} button, the tool packages the trained classifier in an executable code snippet for integration into custom applications, lowering the barrier for the user to program the machine learning-based classifier for deformation recognition. Furthermore, the user has the option to reload the classifier by clicking on the \textit{Preload} button for iterative adjustments. For immediate use without retraining, users may also load our pre-trained evaluation model via Preload when their prints follow our reference FluxIO configuration (\textit{i.e.,} similar structure/lattice density and assembly). The erroneous collection can be cleared by clicking on the \textit{Clear Data} button.

\section{Implementation and Fabrication}
\label{sec:implementation}

\subsection{FluxEditor Implementation}
We implemented FluxEditor in C\# using the RhinoCommon API \footnote{RhinoCommon API: https://developer.rhino3d.com/api/rhinocommon/} and the Grasshopper Human UI\footnote{Grasshopper Human UI: https://www.food4rhino.com/en/app/human-ui}. The design editor runs as a Grasshopper plug-in within Rhino3D 7, a popular CAD software, and utilizes several open-source libraries to support geometry processing and model structuring. Below, we describe how each essential component of the design editor is realized.

\paragraph{SMA Channel Construction.} To house the SMA wire, the editor computes a channel along an approximated medial axis of the user-selected model region. Starting from two clipping planes that are defined by the user and chop the 3D body into a target segment, the editor iteratively slices the volume and finds centroids of the resulting cross-sectional surfaces. These centroids form a smooth trajectory through the model, which the editor uses as the media axis of the selected segment to generate a cylindrical cavity for housing the SMA coil. 

\paragraph{Lattice Generation.} The lattice structures are generated using \textit{Crystallon}\footnote{Crystallon: https://www.food4rhino.com/en/app/crystallon}, a Grasshopper add-on for unit-cell-based modeling. 
Within this library, the gyroid unit cell is given a constant shell thickness of 1~mm and the size of each unit cell is dynamically controlled by a user-facing \textit{Elasticity} slider, which adjusts the perceived stiffness of the structure. 
As presented as ``Elasticity'' for better user interpretability, this input internally maps to a predefined range of lattice solidities (11\%–15\%), where lower solidity values correspond to larger cell sizes and thus more compliant, deformable structures.
The shell thickness and solidity range were empirically determined based on experimental explorations to ensure sufficient structural compliance while supporting reliable SMA-driven deformation.

\paragraph{Parallel Helix Cable Structure Forming.} The outer surface wireframe is constructed from a set of parallel helix cables sampled along the surface. 
The spacing between cables is fixed at 8~mm, a value determined through empirical testing to ensure structural consistency and reliable printability.
Each helix cable is projected onto the model’s surface to conform to its local curvature. The editor converts each resulting helix cable into a tubular strut with a fixed diameter of 1.8~mm, forming a surface-aligned structure that functions both as an anchoring frame and a form-preserving shell for the deformable region.

\paragraph{Anchor Generation.} When \textit{Bending} is selected as the target deformation behavior, the editor automatically generates surface anchoring structures aligned with the user-defined bending direction. These anchors act as passive constraints that guide the deformation along the intended axis. The anchor width is user-configurable through the interface, allowing users to control the degree of constraint and, consequently, the achievable bending angle. 
Although the numerical prediction of the bending angle is not displayed yet, the editor can visualize the expected deformation through the integrated preview (see below), which updates in real time as anchor parameters are adjusted. 

\paragraph{Preview.} The editor displays a displacement-based finite element simulation that estimates the deformation behavior under the selected actuation mode using the \textit{Millipede}\footnote{Millipede: https://www.creativemutation.com/millipede} plug-in. 
With this plug-in, the printed structure is modeled as a linear elastic material with adjustable material parameters (\textit{e.g.,} Young’s modulus and Poisson’s ratio) of the selected material.
To approximate SMA actuation, the contraction forces of the SMA are calculated using the SMA design equations~\cite{SMADesign} and converted into displacement values in the preview. 
This lightweight simulation offers designers a realistic preview of the resulting shape-change behavior before fabrication.

\subsection{FluxShaper Implementation}
We implemented FluxShaper using the P5.js framework for its interface, leveraging ml5.js\footnote{ml5.js: https://ml5js.org/} and TensorFlow.js\footnote{TensorFlow: https://www.tensorflow.org/js} for machine learning integration. To acquire the inductive signal for sensing, we connected a \textit{Texas Instruments LDC1614 evaluation board} to a Python server via serial communication. We bridged this data stream to the browser interface using a Node.js client that securely relays serial port data to WebSocket endpoints, avoiding direct browser access to the hardware.

For deformation recognition, we chose Long Short-Term Memory (LSTM) for its ability to capture temporal dependencies in time-series data~\cite{hochreiter1997long}. It is critical to distinguish the subtle deformation patterns that we set. The model processes input sequences through a sliding window updated every 50~ms to balance latency and prediction stability. To train and evaluate the classifier across varying lattice solidity, we fabricated the same cylindrical lattice geometry at five solidity levels and trained separate LSTM models for each level. We recruited six participants. For each solidity level, we collected 10 trials per participant for each of the seven classes of deformation, yielding 420 labeled sequences per solidity level. We applied dropout (0.3) and L2 regularization ($\lambda=0.001$) to mitigate overfitting, and trained with the Adam optimizer (learning rate = 0.001) for 200 epochs; the validation loss stabilized after around 150 epochs.

The average F1 scores across solidity levels were 88.7\%, 85.6\%, 82.5\%, 80.0\%, and 77.8\% (solidity level from 11\% to 15\% with an interval of 1\%). It decreased monotonically with increasing solidity, consistent with the observation that lower solidity produces larger deformations and thus clearer, more easily distinguishable signals. We reported the confusion matrix with the samples of five solidity levels in the Appendix; it shows that ``Twisting'' has a lower accuracy for recognition, since ``Twisting'' tends to be mislabeled as ``Extension'', ``Compression \& Twisting'', or ``Extension \& Twisting''. Meanwhile, we also found that ``Bending'' is easily mispredicted as ``Compression.'' ``Resting'' reached the highest accuracy.

\subsection{Fabrication of FluxIO-based Models}

To 3D print FluxIO-based models, we follow the conventional SLA printing process using elastic silicone resin (\textit{i.e.,} Formlabs Silicone 40A resin\footnote{Formlabs Silicone 40A Resin: \url{https://formlabs.com/store/materials/silicone-40a-resin/}}) and a consumer-grade SLA 3D printer (\textit{i.e.,} Form 4B\footnote{Form 4B 3D Printer: \url{https://formlabs.com/3d-printers/form-4b}}). 
The 3D-printed objects are cleaned and cured through the standard SLA post-processing steps, including isopropyl alcohol (IPA) rinsing and UV curing.

To create an elastic, lattice-based body that is 3D printable and enables uniform deformation, we aim to determine the minimum wall thickness and the range of unit cell sizes that can be printed without the need for supporting materials. 
As the design specifications of the selected silicone-like material suggest a 1.0~mm wall thickness on the XY plane for printing, we tested gyroid-based lattice units with horizontal wall thickness ranging from 0.6~mm to 1.0~mm, identifying 1.0~mm as the minimum wall thickness that can be reliably printed. 
With the minimum wall thickness fixed, we explored the unit cell size range by creating 21 lattice-based specimens, which used a cylinder with a diameter of 50~mm and the central 32~mm-long portion converted into a lattice-based part, varying the solidity from 5\% to 25\%.
Our experimental results showed that specimens with solidity less than 11\% failed because their cell size (greater than 2~7mm) was too large to print without support. In contrast, specimens with solidity greater than 15\% were too rigid to be activated by the embedded SMA spring. Therefore, models with solidity between 11\% and 15\% were both printable and capable of supporting SMA-driven deformation.
Finally, we also examined the maximum body size that could be actuated by the selected SMA spring before the localized deformation occurred. By testing 14 cylindrical specimens with a diameter ranging from 30~mm to 100~mm (with an increment of 5~mm) without changing the specimen height, we found that, starting with a diameter of 60~mm, the SMA spring could only actuate a limited region around the central axis of the cylinder, making it difficult to deform the areas on the perimeter. 

For the parallel helix cables on the surface, through a series of empirical printing tests with the selected material, we found that a wire thickness of 1.8~mm, a 45\textdegree slope, and an 8~mm helix spacing offered the best balance between printability and elasticity---resulting in a structure that is durable yet flexible under deformation.

\begin{figure*}
    \centering
    \includegraphics[width=0.8\linewidth]{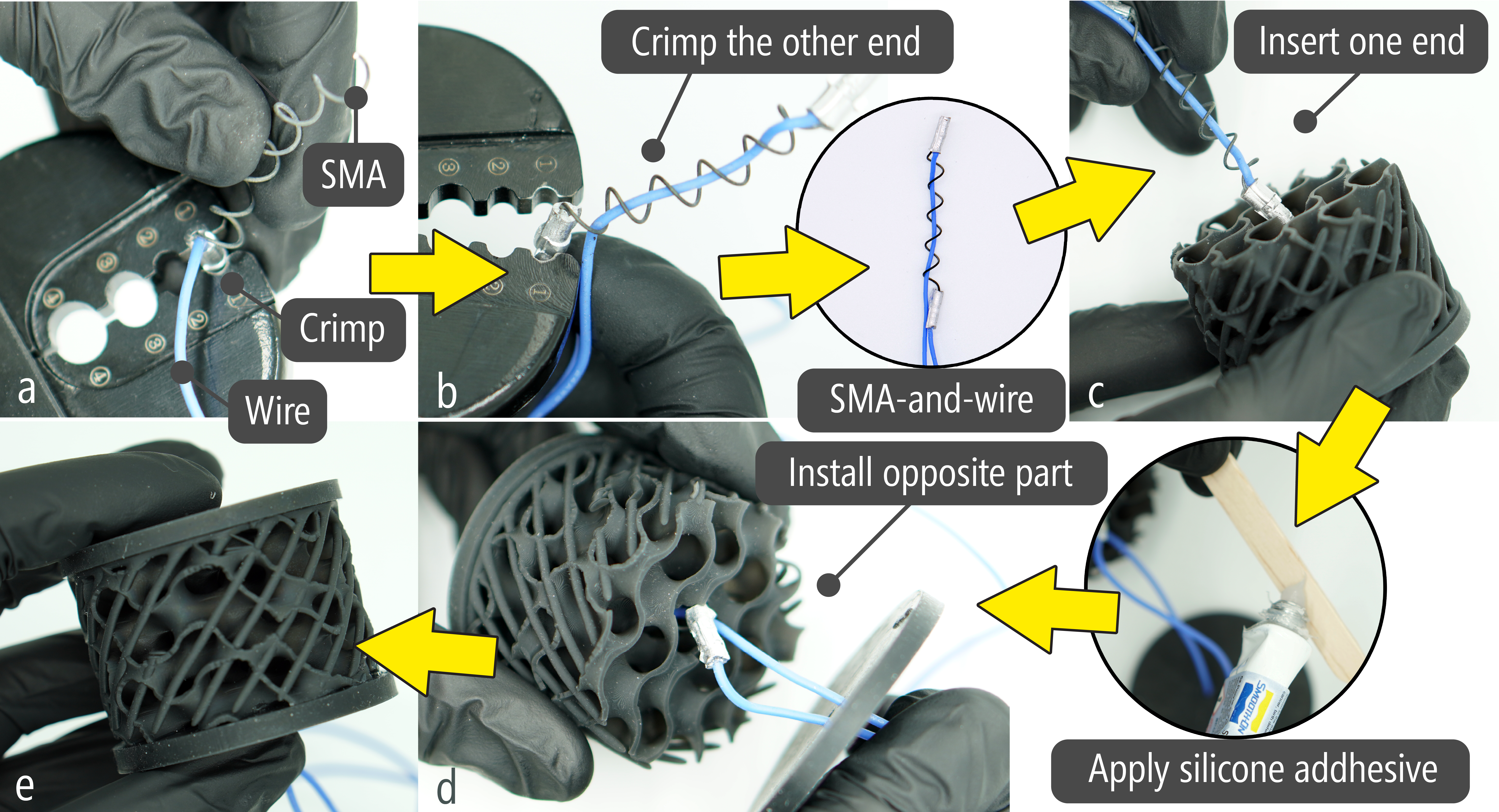}
    \caption{The post-printing process includes: (a) crimp the SMA spring and the conductive wire on one end, (b) crimp another end while the wire goes through the spring's central body, (c) insert one end of the SMA-and-wire group into the model, (d) install the other end of the group with the other part of the model, and (e) bond two model parts with silicone adhesive.}
    \Description{This figure shows the post-print processes }
    \label{fig:postprint}
    \vspace{-2mm}
\end{figure*}

\subsection{SMA Spring Installation}

To prepare the SMA actuator, we reserve a cylindrical channel with a diameter of 10~mm (the selected commercial two-way Nitinol SMA spring\footnote{NexMetal Nitinol spring shape memory alloy: https://nexmetal.com/} has a diameter of 7~mm and a 3~mm gap) to house the embedded SMA spring and conductive wire in the center. We cut the SMA spring to approximately match the length of the printed channel and remove any end loops commonly found in off-the-shelf SMA springs. A 1.2~mm crimp and 24~AWG wire are used to connect one end of the spring (Fig.~\ref{fig:postprint}a), with the wire threaded through the center of the coil to reduce strain during contraction (Fig.~\ref{fig:postprint}b). The opposite end is extended using a second crimp and wire in the same manner.
The printed SMA channel includes an integrated socket design to secure the SMA-and-wire assembly (Fig.~\ref{fig:postprint}c\&d). Each socket features a hemispherical hook that prevents the crimped ends from slipping out during actuation. The SMA assembly is inserted manually, allowing reuse or replacement.
The bottom section of the model is printed separately to facilitate easier insertion of the SMA assembly (Fig.~\ref{fig:postprint}d). After insertion, it is attached to the main body using silicone epoxy adhesive (Fig.~\ref{fig:postprint}e).

\subsection{Actuation Control}
The actuation of the SMA spring is achieved through Joule heating, where an electrical current is passed through the alloy to raise its temperature above the austenite start temperature (45~\textdegree C), inducing contraction and driving the desired deformation in the FluxIO structure. 
To ensure precise and safe control, we employ a pulse-width modulation (PWM) scheme, which allows for adjustable heating rates and prevents overheating that could be dangerous or degrade the SMA's performance over repeated cycles.

The control circuit consists of a microcontroller interfaced with a power MOSFET to handle the SMA's current requirements, up to 1~Amp at 5~V, as determined through empirical testing for reliable actuation. 
The SMA is connected in series to the MOSFET’s drain, and the gate is driven by a PWM signal from the microcontroller. 
A current-limiting resistor is included to stabilize current flow and protect against overdraw, ensuring consistent deformation amplitudes while minimizing thermal stress on the surrounding lattice.

To enable both actuation and inductive sensing using the same SMA coil, a switching circuit is incorporated to alternate between the two modes, preventing interference between the heating current and the inductance measurement. 
The switching is achieved using a pair of single-pole double-throw (SPDT) relays controlled by the microcontroller. 
In actuation mode, the relays connect the SMA terminals to the MOSFET and power supply; in sensing mode, they reroute the SMA to the connected LDC1614 evaluation board. 
The microcontroller coordinates mode transitions via digital output pins, ensuring sensing occurs only during non-actuation periods, such as cooling phases or idle states. 
With a switching time of less than 10~ms, this time-multiplexed approach enables near-real-time deformation recognition while preserving actuation functionality.
This time-multiplexed approach balances the dual functionality of the SMA, as it acts both as an actuator and a sensor.
For example, upon detecting a \quotes{Bending} input, the system switches to actuation mode and initiates a PWM cycle to contract the SMA. 
Actuation typically requires 2–5 seconds for full contraction, followed by a cooling period of 10–20 seconds under ambient conditions or assisted by passive airflow, during which sensing resumes. The actuation and recovery periods may vary when the SMA is embedded into different FluxIO-based mechanical designs.

\section{Example FluxLab Applications}
\label{sec:app}
To demonstrate the potential of our system for prototyping sensing and actuation capabilities in various applications, we built three example prototypes that showcase a self-deformable steamer bowl clip, a remotely controlled gripper, and a dinosaur-shaped smart desk lamp for kids. 

\subsection{Self-deformable Steamer Bowl Clip}
\begin{figure}[h]
    \centering
    \includegraphics[width=1\linewidth]{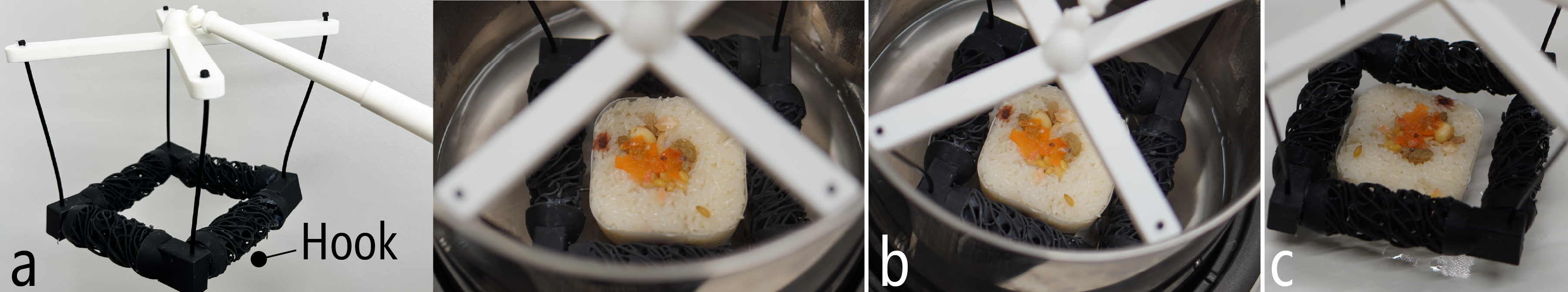}
    \caption{The self-deformable steamer bowl clip (a) converts a four-bar hook at the bottom into FluxIO-based mechanisms so the bars (b) contract to grip the rim of the hot bowl when the clip is moved in a hot steamer. After the bowl is picked and moved outside the steamer and the bars cool down, (c) the user can release the hook from the bowl. }
    \Description{The figure shows the self-deformable steamer bowl clip. }
    \label{fig:app1}
    \vspace{-2mm}
\end{figure}

Steamer bowl clips are commonly used to safely lift hot bowls, plates, or trays from steamers.
We created a shape-changing steamer bowl clip using a four-bar square hook with four corners hanging on an extended handler (Fig.~\ref{fig:app1}a).
The four edges of the hook were converted into FluxIO-based mechanisms with FluxEditor, retaining the wavy surfaces to provide sufficient friction to grip the rim of the hot bowl (Fig.~\ref{fig:app1}b).
To use this modified clip, the user first places the clip in a hot steamer and lowers it until the hook is positioned underneath the rim of the hot bowl. 
The hot steam heats the SMAs embedded in the hook bars, driving them to compress and thus shortening the four bars (Fig.~\ref{fig:app1}c).
Once the hook becomes smaller, the user lifts the hot plate with the bars firmly gripping the rim of the bowl and places the bowl on a table.
After the clip cools down to room temperature, the bars elongate again, releasing the clip from grabbing the bowl. 
With a long handle, this self-deformable, circuit-free steamer bowl clip protects the user's hand from burns.

\subsection{Remotely Controllable Robotic Gripper}
\begin{figure}[h]
    \centering
    \includegraphics[width=1\linewidth]{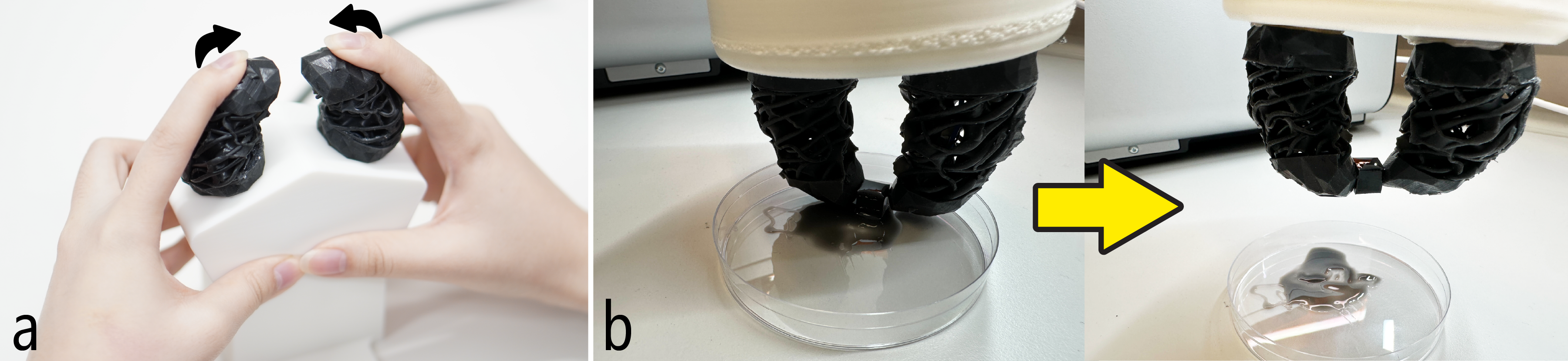}
    \caption{The user (a) bends a pair of gripper models at home to (b) remotely control twin grippers in the lab to pick up an uncured fresh print.}
    \Description{The figure shows the remotely controllable gripper application. }
    \label{fig:app2}
    \vspace{-2mm}
\end{figure}

To protect an experimenter from touching the uncured resin when picking up a freshly printed object, we created a pair of remotely controllable grippers (Fig.~\ref{fig:app2}).
The grippers were created by converting octopus tentacles into anchor-applied FluxIO units that can laterally bend when heated. 
The user can remotely control the bending behaviors of each gripper by bending a gripper twin.
The gripper twin senses the bending input by the user and transmits the control message to the corresponding operating gripper, executing the bending movement by using a connected circuitry to heat the integrated SMA.
By manipulating the two gripper twins, the user can remotely control the grasping and releasing of the grippers, enabling real-time object-grasping tasks, such as picking up a fresh resin print. 

\subsection{Smart Desk Lamp}
\begin{figure}[h]
    \centering
    \includegraphics[width=1\linewidth]{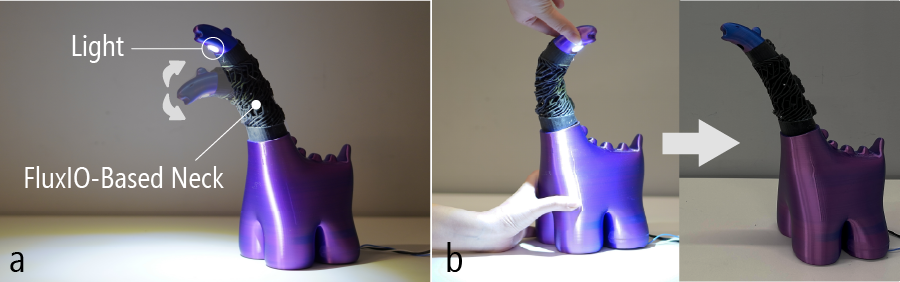}
    \caption{The dinosaur-shaped desk lamp (a) nods its head to remind the kid to take a break, and then the kid can (b) turn off the light by twisting the dinosaur's neck. }
    \Description{The figure shows the dinosaur-shaped smart desk lamp. }
    \label{fig:app3}
    \vspace{-2mm}
\end{figure}

Another example that combines sensing and actuation into one uniform device is the smart desk lamp for children.
We created a long-necked dinosaur-shaped desk lamp with FluxLab and embedded the FluxIO unit in the long neck of the dinosaur.
The anchor was added to the backside of the neck, so the dinosaur's head would sag when the embedded SMA was heated and rise again when the SMA was cooled down. 
An external circuit was used to control the heating.
A bright LED light was installed in the dinosaur's head, and the wires went through the anchor to the body, connected with the controlling circuit as well.
After the child studies for a long period of time, the dinosaur lamp nods its head to indicate the time for a break (Fig.~\ref{fig:app3}a). 
The child can turn off the light by manually twisting the dinosaur's neck (Fig.~\ref{fig:app3}b).

\section{Limitations and Future Work}
\label{sec:discussion}

FluxLab features a core design method---FluxIO---that converts a 3D-printed shape into a shape-changable device with integrated deformation sensing capability by embedding an SMA spring, whose behavior is determined by the printed channel geometry, lattice, and surface anchoring configurations. 
By formalizing this structural coupling as a generalizable design method, FluxLab shifts the focus from assembling discrete components to programming material behavior through structural design, showing how deformation, sensing, and form can be co-designed within a unified workflow. 
This approach provides a hybrid lens for encouraging makers and researchers to approach the making of interactivity, such as sensing and actuation, through structural design and material property control.

While FluxLab demonstrates multifaceted support for experienced makers and researchers to design, fabricate, and control the actuation and sensing capabilities of 3D-printed devices, we outline the limitations of this approach and the next steps in this research below.

\subsection{Limitations}
\subsubsection{Geometric Constraints}
\label{subsubsec:geometric_constraints}
FluxLab currently supports embedding SMA-driven FluxIO units into only 3D models of moderate size and curvature. The design is limited by the minimum SMA channel diameter (\textit{i.e.,} 10~mm) and the smallest printable lattice cell size required to maintain elasticity. Moreover, embedding straight SMA springs into highly curved geometries remains challenging. These geometric constraints make the current approach less suitable for miniature or continuous, high-curvature shapes.

\subsubsection{Fabrication and Material Constraints}
\label{subsubsec:fab_and_material_constraints}
The three-layer FluxIO design method uses elastic, low-hardness materials that enable structural flexibility while maintaining form.
While the selected silicone-like printing material offers great elasticity, it requires a specialized 3D printer (\textit{i.e.,} Form 3 or a more advanced printer model) and fabrication process (\textit{i.e.,} SLA 3D printing and specific solvent for cleaning the uncured silicone resin), which exhibits various fabrication and material challenges.
First, silicone resin is highly dependent on print orientation for surface quality, dimensional accuracy, and mechanical properties.
Since we aim to print the object without support, it takes trials and errors to identify an optimal printing orientation to avoid failed prints.
Second, the material's high viscosity makes it difficult to print elastic structures, such as the lattice structures in our design. 
Each printer layer is subjected to significant tension as it is peeled from the resin tank, which increases the risk of layer tearing during the printing process.
To address this, we manually added thick support structures and carefully arranged them alongside the target model, as well as a delicate post-print support removal process.
Finally, the material imposes strict constraints, such as the minimum wall thickness, for reliable printing, which limits the material and mechanical design space that our approach could otherwise explore. 

\subsubsection{Sensing Performance and Actuation Control}
\label{subsubsec:sensing_actuation_performance}
Our work aims to demonstrate a full suite of design methods and tools to support the creation of 3D printable shape-changing devices with integrated deformation sensing capabilities: the three-layer design method, the design editor, and the deformation sensing authoring tool. While the sensing and actuation capabilities were showcased in the example applications, a systematic evaluation of sensing accuracy and actuation performance has not yet been conducted. As an enabling system, we envision that FluxLab can be used for various applications, with custom shapes being created and bespoke deformation inputs being trained and recognized. The current system demonstrates feasibility but lacks quantitative analysis of sensing resolution, deformation reliability, and actuation force.

\subsection{Future Work}
\subsubsection{Multiple Parts and Design Support}
In the example applications, each printed part contains a single FluxIO structure to demonstrate an individual basic shape change. However, multiple FluxIO structures can be integrated within a single object to support more complex and expressive deformations. For example, connecting FluxIO structures in series allows asynchronous deformations along a segment, while arranging them in parallel enables localized or distributed changes across a larger surface, as implied by our tests in Section \ref{sec:sensor_actuator}. To support these advanced deformation behaviors, FluxLab needs to be extended with additional features, such as spatial layout management and coordinated deformation planning, allowing designers to compose multiple FluxIO elements within a single object.

\subsubsection{Compact and Pre-shaped SMA Integration}
As discussed in Section~\ref{subsubsec:geometric_constraints}, the current design is limited by geometric constraints, such as the difficulty of embedding straight SMA springs into small or highly curved geometries. To address this, future work will explore compact SMA alternatives that provide sufficient actuation force while fitting within smaller structural channels, such as \textit{BioMetal BMX15020}\footnote{BioMetal Helix BMX15020: https://www.biometal.biz/product-page/bmx15020}. We also plan to develop a setup for pre-shaping SMAs into customized geometries, such as curved or S-shaped forms, enabling FluxIO to accommodate a wider range of object sizes and deformation topologies.

\subsubsection{Material and Fabrication Exploration}
The current design uses elastic silicone-like resin that requires specialized SLA printing and is sensitive to printing orientation and material viscosity. As we mentioned in ~\ref{subsubsec:fab_and_material_constraints}, these dependencies limit the reliability of printing and constrain the achievable geometries. Therefore, future work will explore softer and lower-viscosity materials and investigate alternative fabrication methods.
While the currently used material can achieve a Shore A hardness of 40---more elastic compared to other commercially available elastic resin, it is still worth exploring alternative experimental soft materials and fabrication techniques, such as Soft ToughRubber™---a soft DLP printing material with a Shore A hardness of 28.6.
Such explorations could expand the mechanical and aesthetic design space of FluxIO, making the fabrication process more accessible to designers and makers.

\subsubsection{Sensing and Actuation Evaluation}
Future work should also quantify the number of distinct deformation inputs that can be reliably sensed and assess the sensing accuracy across different magnitudes and locations. It is equally important to evaluate the actuation capability by measuring the output force yielded by the FluxIO-based structures with various SMA sizes, as larger SMAs can generate greater force but require a correspondingly larger host structure. Investigating these trade-offs will help make design choices in FluxLab and expand FluxIO's applicability.

\subsubsection{Tool Usability and Accessibility for Non-expert Users}
Currently, FluxLab targets experienced makers and researchers who are familiar with SLA 3D printing and electronic control. We acknowledge that the usability and learning experience of FluxEditor and FluxShaper have not yet been formally evaluated. 
As part of the future work, we will conduct user studies with non-expert makers to assess the overall user experience with the tools and to understand the challenges with the proposed approach, such as how easily they can master the tools and how reliable and effective these tools are for users to achieve their intended deformation behaviors with the created 3D designs.


\section{Conclusion}
\label{sec:conclusion}
We presented FluxLab, a system that enables the design and fabrication of 3D-printed shape-changing devices with integrated deformation sensing. 
We introduced a three-layer structural design method, FluxIO, which comprises a central SMA channel for sensing and actuation, gyroid-based lattice cells serving as padding structures in the middle to support structural elasticity, and a parallel helix-based surface to preserve overall aesthetics. 
To enable users to convert 3D models into FluxIO-based bodies, we developed a design editor, FluxEditor, that converts models into 3D-printable bodies with silicon resin on a commercialized SLA 3D printer with minimal post-printing processing. 
To enable users to apply deformation sensing to the printed devices, we also developed a deformation authoring tool, FluxShaper, which allows for the creation and export of a machine learning-based classifier that distinguishes desired deformation behaviors using inductive sensing. 
To demonstrate our approach, we built three applications with FluxLab, including a self-deformable steamer bowl clip, a pair of remotely controllable grippers, and an interactive desk lamp for children.


\bibliographystyle{ACM-Reference-Format}
\bibliography{reference}

\appendix
\section{Appendix: Evaluation Results}
\begin{figure}[H]
  \centering
  \includegraphics[width=\columnwidth]{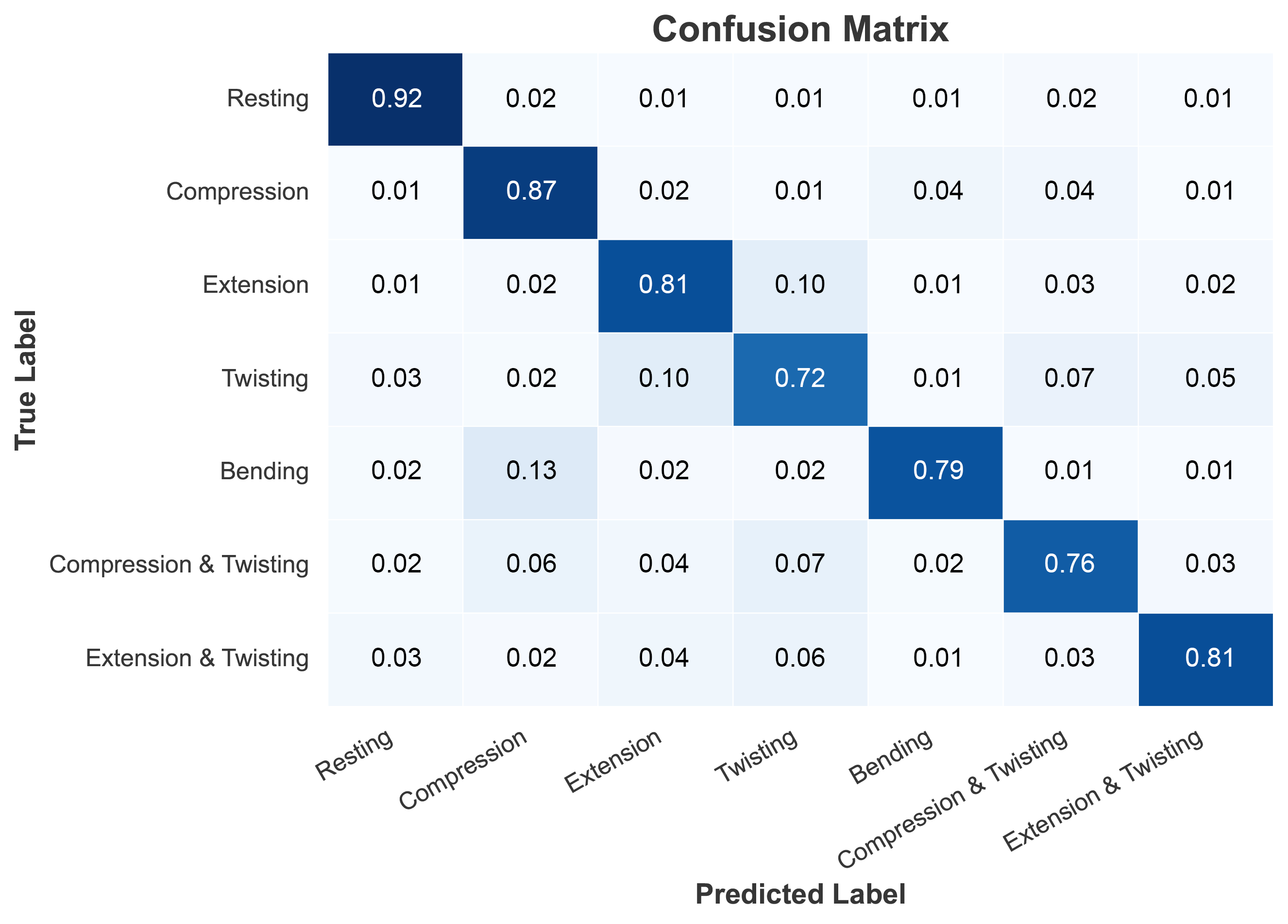}
  \caption{Averaged confusion matrix of the deformation classifications for the five cylindrical samples.}
  \label{fig:confmatrix}
\end{figure}

\end{document}